\documentclass[
    aps,
superscriptaddress,
    prl, 
    twocolumn,
    letterpaper,
    10pt,
    floatfix
]{revtex4-1}

\usepackage{enumerate,graphicx,amsmath,amssymb}
\usepackage{epstopdf}
\usepackage{siunitx}
\usepackage{color}

\def\beq{\begin{equation}}
\def\eeq{\end{equation}}
\newcommand{\delete}[1]{}

\newcommand{\be}{\begin{equation}}
\newcommand{\ee}{\end{equation}}

\begin{document}

\title{Searching for ultra-light dark matter with optical cavities}

\author{Andrew A. Geraci}
\affiliation{Department of Physics and Astronomy, Northwestern University, Evanston, Illinois 60208, USA}

\author{Colin Bradley}
\affiliation{Department of Physics, University of Nevada, Reno, Nevada
89557, USA}
\author{Dongfeng Gao}
\affiliation{State Key Laboratory of Magnetic Resonance and Atomic and Molecular Physics, Wuhan Institute of Physics and Mathematics, Chinese Academy of Sciences, Wuhan 430071, China}

\author{Jonathan Weinstein}
\affiliation{Department of Physics, University of Nevada, Reno, Nevada 89557, USA}

\author{Andrei Derevianko}
\affiliation{Department of Physics, University of Nevada, Reno, Nevada 89557, USA}

\begin{abstract}
We discuss the use of optical cavities as tools to search for dark matter (DM) composed of virialized ultra-light fields (VULFs). Such fields could lead to oscillating fundamental constants, resulting in oscillations of the length of rigid bodies. We propose searching for these effects via differential strain measurement of rigid and suspended-mirror cavities. We estimate that more than two orders of magnitude of unexplored phase space for VULF DM couplings can be probed at VULF Compton frequencies in the audible range of 0.1-10 kHz.
\end{abstract}
\maketitle

{\em Introduction---} 
Despite overwhelming observational evidence for the existence of dark matter (DM), its composition and non-gravitational interaction with standard model fields and particles remain a mystery~\cite{Feng2010,Bertone2005,Profumo2017-Book}. Assuming DM is composed of new particles or fields, the mass of the DM constituents can vary over a vast range of fifty orders of magnitude, see, e.g.,~\cite{DerPos14}.  Most particle-physics experiments so far have searched for weakly interacting massive particle (WIMP) DM with masses at the $\sim \mathrm{GeV}$ scale. With no uncontested evidence for such DM to date, and a broad range of possible DM candidates, it is worthwhile to broaden the toolbox of techniques to explore other mass scales and couplings.

Among viable candidates are ultra-light fields with masses below $\sim$ 10 eV, which behave as classical fields rather than individual particles (see recent review \cite{Safronova2017-RMP}).  As such, they require  precision measurement techniques rather than direct observations of energy deposition in particle detectors. A well-motivated example of such a light field is the QCD axion;  axion-like or other scalar or vector (e.g., dark-photon) DM candidates have also been proposed. We collectively refer to such
candidates as VULFs (virialized ultra-light fields).
A variety of experimental techniques have been used or proposed for VULF searches,  including resonant cavities, torsion balances, atom interferometers, atomic clocks, molecular absorption, and magnetometers~\cite{ArvHuaTil15,GraKapMar2016,GeraciDerevianko2016-DM.AI,casper,StaFla2015-LaserInterferometry,StaFla2015,Arvanitaki2016_SoundDM,Arvanitaki2016-GWdetectors,BraClaKin03}.

One of the viable VULF candidates is a scalar field, motivated by string theory dilatons and moduli~\cite{Dimopoulos1996,Arkani-Hamed-2000,Burgess2011,Cicoli2011,Taylor1988,DamourPolyakov1994}. Such fields  lead to apparent variation of fundamental constants such as the fine-structure constant $\alpha$ or the mass of the electron $m_e$~\cite{DamourDonoghue2010}. On timescales short compared to the VULF coherence time, the DM field can be expressed as
    \begin{equation}
    \phi(t,\mathbf{r}) \approx \frac{ \hbar}{m_\phi c} \sqrt{2 \rho_{\rm{DM}}}
     \cos{[ 2 \pi f_\phi t-\mathbf{k_\phi}\cdot \mathbf{r} + ...]} \, ,
\end{equation}
where $\rho_{\rm{DM}}\approx 0.4 \, \mathrm{GeV}/\mathrm{cm}^3$ is the local DM energy density,  $m_\phi$ is the mass of the DM field,
$f_\phi= m_\phi c^2/(2 \pi \hbar)$ is Compton frequency,
and $k_\phi = m_\phi v/\hbar$ with $v$ being the velocity of  DM with respect to the instrument.  Detailed discussion of the expected coherence properties of VULFs can be found in Ref.~\cite{Derevianko2016a}.

In the dilaton-like models, VULFs drive oscillations of the electron mass and fine structure constant,
\begin{eqnarray}
\frac{\delta m_e(t,\mathbf{r})}{m_{e,0}} =d_{m_e}\sqrt{4\pi \hbar c} E_P^{-1} \phi(t,\mathbf{r}) \, , \label{Eq:mass-variation} \\
\frac{\delta \alpha(t,\mathbf{r})}{\alpha_0} =d_{e}\sqrt{4\pi \hbar c} E_P^{-1} \phi(t,\mathbf{r}) \, \label{Eq:alpha-variation}.
\end{eqnarray}
Here $d_{m_e}$ and $d_e$ are dimensionless dilaton couplings and $E_P \equiv \sqrt{\hbar c^5/G}$ is the Planck energy.
These effects could be searched for with atomic clocks and interferometers~\cite{ArvHuaTil15,GeraciDerevianko2016-DM.AI}, however, due to the finite interrogation time, they are limited to Compton frequencies of order 1 Hz and below. At higher frequencies, DM-induced strain in solids is a promising approach.
The DM-induced oscillations (\ref{Eq:mass-variation},\ref{Eq:alpha-variation}) of the fine structure constant and electron mass drive oscillations in the Bohr radius $a_{0}=\hbar/(\alpha m_{e}c)$ and, therefore, in the size of atoms and chemical bonds. For sufficiently slow oscillation frequencies, this causes a time-varying strain $h$ in solid materials, given by $h=-\frac{\delta\alpha}{\alpha_0}-\frac{\delta m_{e}}{m_{e,0}}
$, where we have ignored small relativistic effects.

Resonant bar detectors have been previously proposed as a method of detecting  ultra-light-DM-induced strain in material bodies~\cite{Arvanitaki2016_SoundDM}. This approach relies on the resonant enhancement of the vibration of the bar relative to the surrounding objects, in order to differentiate the DM signal from the expansion and contraction of the remainder of the apparatus. Consequently, that approach is inherently a resonant method, with sensitivity significantly degraded off-resonance. Here, we propose using two optical cavities --- with different sensitivities to DM-induced strain --- to search for the same signal. In contrast to resonant bar detectors, the method we propose is {\em broadband}. Despite the lack of resonant enhancement, the Allan deviation of a laser locked to an optical cavity --- superior to all other current technology at  times $\lesssim 1$~s --- is anticipated to extend the discovery reach for ultralight scalar field DM by up to 3 orders of magnitude in the 0.1-10 kHz frequency band, corresponding to $m_\phi \approx 10^{-13} - 10^{-11} \,\mathrm{eV}/c^2$.  Our method thus closes a gap in the mass range for VULF searches in the audible frequency band.

\emph{Proposed experiment---}
We consider an arrangement of two co-located  high-finesse Fabry-Perot optical cavities. The first optical cavity is constructed with mirrors connected by a rigid cavity spacer, as is typical for optical clocks. The second optical cavity consists of two mirrors suspended by pendulums, as is used in LIGO, with a resonant mechanical frequency below the frequencies of interest. This suppresses the sensitivity of the second cavity's length to high-frequency variations in the length of its supporting spacer.  Thus, if the size of atoms oscillates in time, the length (and hence resonant frequency) of the first cavity should oscillate with respect to the second. The experimental technique described below essentially measures differential strain of the two cavities. The VULF signal would appear as a spike in the power spectral density (PSD) of the measured differential strain. The VULF signal is predicted~\cite{Derevianko2016a} to exhibit a strongly asymmetric profile of width $\Delta f_\phi \approx 3 \times 10^{-6} f_\phi$. This distinct signature should allow 
to discriminate the VULF signal from many conventional noise sources.

\begin{figure}[!t]
\begin{center}
\includegraphics[width=0.9\columnwidth]{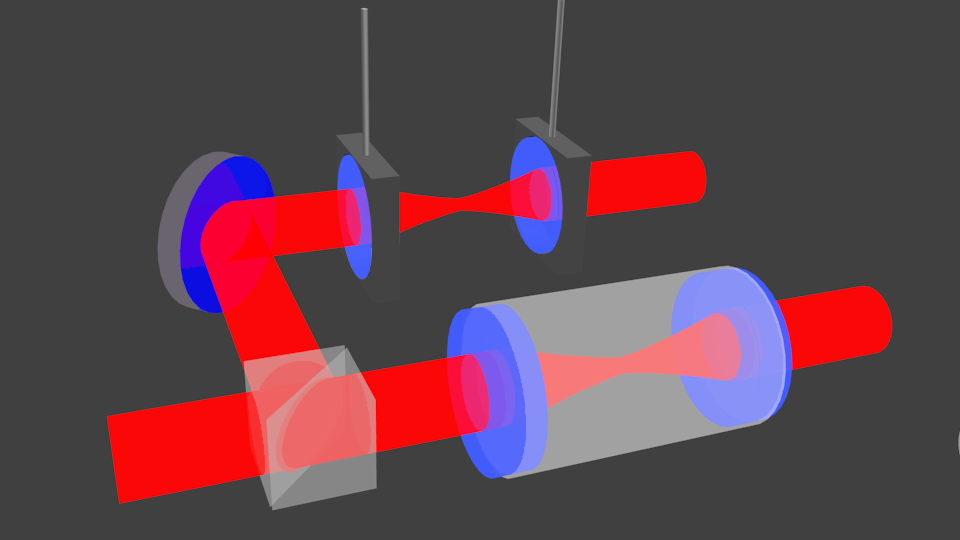}
\caption{Experimental setup. Light from a common laser is directed into two Fabry-Perot resonators, one with suspended mirror substrates and one with a rigid cavity spacer. Modulation in e.g. the electron mass due to dilatonic DM at the 0.1-10 kHz frequency range results in periodic length changes in the rigid cavity, while the DM-induced length changes in the suspended interferometer are suppressed by the low frequency mechanical suspension. 
\label{fig:apparatus}}
\end{center}
\end{figure}

As shown in Fig.~\ref{fig:apparatus}, light from a single laser is sent into both cavities. The cavities are located on a single optical table to suppress differential Doppler shifts of the laser light due to relative cavity motion. The laser frequency is locked to the first cavity using Pound-Drever-Hall (PDH) locking. PDH provides a high stabilization bandwidth not limited by the cavity response time \cite{drever1983laser}. The light traveling to the second cavity is frequency shifted onto resonance with the second cavity using an acousto-optic modulator (AOM). The frequency of the AOM is modulated to lock its transmitted light to the resonance of the second cavity using PDH. The AOM drive frequency is recorded directly (or mixed down to a lower frequency for lower-bandwidth recording), which provides the frequency shift $\Delta_f(t)$ of the resonant frequency of one cavity relative to the other as a function of time.
The strain of one cavity relative to the other is simply $h(t) = \Delta_f(t)/f_0$, where $f_0$ is the nominal frequency of the laser. We consider three possible cavity lengths, of 10 cm, 30 cm, and 100 cm in order to provide coverage over the audible frequency band. While all cavities are broadband in detection, the choice of cavity length is a trade-off between strain sensitivity and maximum detectable frequency, as discussed below. The proposed experimental parameters are shown in Table I taking the 30 cm cavities as an example.

The minimum detectable differential strain is limited by photon shot noise, as calculated below. However, the differential strain itself can originate not only from VULF DM, but also in technical noise sources, such as thermal fluctuations of the cavity spacers, the mirrors, and the mirror coatings. The limits which can be placed on 
VULF DM couplings are limited by these fluctuations as discussed below.



\begin{table}[!t]
\small
\begin{center}

   \begin{tabular}{@{}ccc@{}}
  \hline
  \hline
  Parameter & Value & Description \\
  \hline
$L$ & 30 cm & Cavity spacer length \\
$\mathcal{F}$ & $10^4$ & Cavity finesse  \\
$w$ & 2 mm & Laser waist\\
$\lambda$ & 1550 nm & Laser wavelength\\
$P$ & 1 mW & Incident laser power \\
$t$ & 2 cm & Mirror substrate thickness  \\
$r$ & 3 cm & Mirror substrate radius  \\
$d$ & $4 \times 10^{-6}$ m  & Coating thickness \\
$\phi_c$ & $2.7 \times 10^{-4}$ & Loss angle coating \\
$\Phi$ & $10^{-7}$ & Loss angle mirror substrate \\
$\phi_{sp}$ & $10^{-6}$ & Loss angle spacer \\
$\phi_{susp}$ & $2 \times 10^{-7}$ & Loss angle suspension ( $> 1$ kHz) \\
$d_{\rm{wire}}$ & 310 $\mu$m & Suspension wire diameter \\
$L_{\rm{wire}}$ & 8 cm & Suspension wire length \\
$R_{sp}$ & 3 cm & Outer radius spacer \\
$r_{sp}$ & 0.5 cm & Inner radius spacer \\
$Y$ & 70 GPa &  Young's modulus, substrate and spacer \\
$\sigma$ & 0.25 &  Poisson ratio, substrate and spacer \\
$Y_c$ & 110 GPa &  Young's modulus, coating \\
$\sigma_c$ & 0.22 &  Poisson ratio, coating \\
$T$ & 300 K &  Cavity temperature\\
  \hline
  \hline
  \end{tabular}

\caption{\label{table1} Experimental parameters chosen for the cavity with rigid spacer and suspended mirror cavity.  }
\end{center}
\end{table}

\begin{figure}[!t]
\begin{center}
\includegraphics[width=1.0\columnwidth]{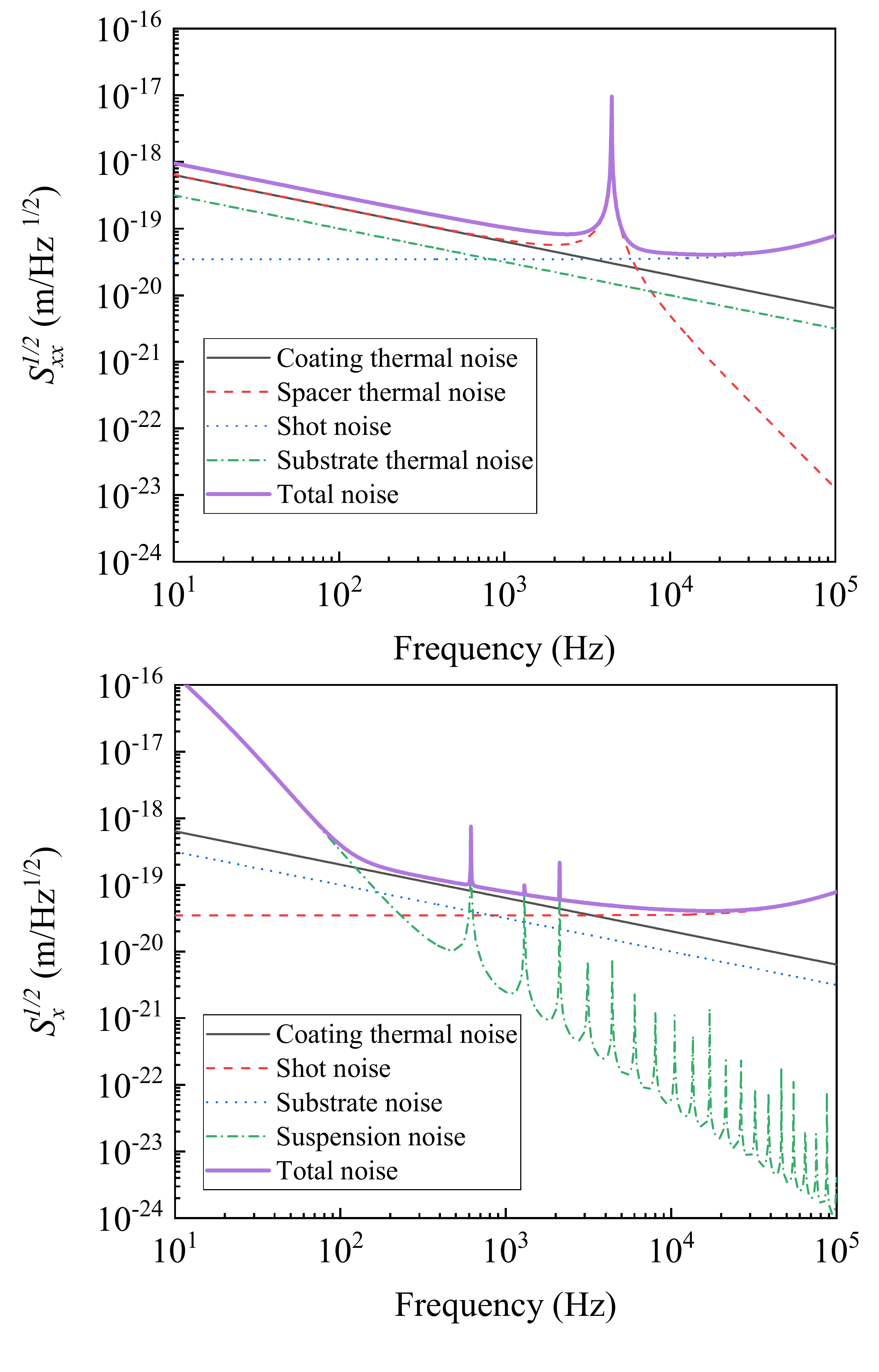}
\caption{Expected noise sources for the proposed experimental geometry, with parameters of Table~\ref{table1}. Upper panel: cavity with rigid spacer. Lower panel: cavity with suspended mirror substrates.
\label{Fig:Noise}}
\end{center}
\end{figure}


\emph{Noise sources and systematic effects.}
There are several fundamental and technical sources of noise which limit the ability to measure the effective strain. A list of fundamental and technical noise estimates are included in Fig.~\ref{Fig:Noise} for the geometry we consider.
In quantifying various cavity noise sources, we operate in terms of one-sided displacement PSD $S_{xx}(f)$ related to the differential strain PSD as 
$S_{hh}(f) = S_{xx}(f)/L^2$, where $L$ is the cavity length.
The dominant sources of thermo-mechanical noise due to intrinsic dissipation tend to improve at higher frequency $f$ as $f^{-1/2}$. Thus at frequencies above 10 kHz it is possible to realize shot-noise limited position detection. In the future, squeezed light offers the prospect for further improved sensitivity. 

Photon shot noise limits the minimum detectable phase shift to $\delta \phi \sim  1/(2\sqrt{I}) \sqrt{b}$, where $I=P/\hbar\omega_c$ is the incident photon flux from a laser of power $P$ and frequency $\omega_c$ and $b$ is the measurement bandwidth \cite{hadjar}. The corresponding photon shot-noise limited one-sided displacement PSD is $ 
S_{xx}(f)=S_{xx,0}[1+(2 \pi f)/\Omega_{cav}],$ for an impedance matched cavity of line-width $\Omega_{cav}$ \cite{hadjar}.  Here $S_{xx,0}=\frac{\lambda}{16 \mathcal{F} \sqrt{I}}$, with 
$\mathcal{F}$ being the cavity finesse and $\lambda$ the laser wavelength.

Coating thermal noise is a common limitation in precision optical cavity metrology both for the optical clock community and the laser interferometer gravitational wave observatories \cite{coatingnoisesuspended}. Assuming similar coating and substrate mechanical parameters, we can arrive at a simplified expression for the noise \cite{coatingnoisesuspended, chalermsongsak2014broadband}

\begin{equation}
S_{xx,coat}(f)=\frac{2 k_B T}{\pi^{3/2} f}\frac{1}{Y_c w (1-\sigma_c^2)} \phi_{coat} \, ,
\end{equation}
where $\phi_{coat}=\frac{2d}{\pi^{1/2}w} \frac{(1-2\sigma_c)}{(1-\sigma_c)}\phi_{c},$
for a coating at temperature $T$ with Young's modulus $Y_c$, Poisson ratio $\sigma_c$, beam waist $w$, and coating loss angle parameter $\phi_c$. Here, as in Ref.~\cite{chalermsongsak2014broadband}, we are assuming for simplicity that the coating properties are isotropic and we choose a multilayer dielectric stack of materials adding to thickness $d$ and also assume a similar effective loss angle as in~\cite{chalermsongsak2014broadband}. Our estimated coating noise level is about an order of magnitude less than that demonstrated in~\cite{chalermsongsak2014broadband} because our chosen beam waist is nearly ten times larger and we expect that our estimate is reasonable given similar materials. 

Spacer thermal noise in the rigid cavity contributes at a level similar to the mirror coatings for the experimental parameters considered. A simple harmonic oscillator in the low frequency regime has a spectral density \cite{saulson}
$$S_{xx,sp}(f)=\frac{4k_BTk\phi_{\rm{sp}}}{(2 \pi f)[(k-m(2 \pi f)^2)^2+k^2\phi_{\rm{sp}}^2]} \,,$$
with $k_B$ being Boltzmann's constant, $T$ being the temperature, $k$ being the effective spring constant, and $\phi_{\rm{sp}}$ being the loss factor of the material. Provided that the cylinder is homogeneous, the approximation of a simple harmonic oscillator is adequate. This assumption also yields the effective spring constant: $k=\frac{YA}{L}$ where $Y$ is the Young's modulus of the material, $A$ is the cross sectional area, and $L$ is the length of the rod. For longitudinally driven oscillations near resonance, the effective mass is half the mass of the cylinder. The resulting spectral density is given in Fig.~\ref{Fig:Noise} which uses values for material properties from Refs.~\cite{mat1,chalermsongsak2014broadband}. These results could improve significantly with the use of synthetic fused silica, as the loss factors can be much lower.
In the low-frequency limit, for the geometry of a cylinder of radius $R_{\rm{sp}}$ with a hole of radius $r_{\rm{sp}}$ bored through the center, the position spectral density due to thermal noise is given by the form\cite{chalermsongsak2014broadband}
\begin{equation}
S_{xx,sp}(f)=\frac{4k_B T}{\pi f} \frac{L}{2\pi Y (R_{\rm{sp}}^2-r_{\rm{sp}}^2)}\phi_{sp} \, .
\end{equation}

For the cavity which consists of freely suspended mirrors, the thermal noise in the wire suspension contributes in a manner similar to that in the LIGO type detectors. We choose a fused silica wire suspension of length $L_{\rm{wire}}$, diameter $d_{\rm{wire}}$, and effective loss angle $\phi_{\rm{susp}}$ as specified in Table~\ref{table1}, and study the pendulum, torsion, and violin modes assuming a simple single-wire suspension and modal approximation \cite{ligosusp}. For the loss factor, we take a frequency dependent model as given in Ref. \cite{ligosusp} which asymptotes to the value in Table \ref{table1} above $\sim 1$ kHz. Apart from a few discrete narrow peaks corresponding to the violin mode frequencies, we expect the suspension noise to be sub-dominant to other thermal noise sources. For a realistic design, a tapered wire diameter may be chosen to further improve losses, as in LIGO \cite{ligosusp}.

The thermal Brownian motion of the mirror substrates takes the form \cite{mat1}
\begin{equation}
S_{xx,sub}=\frac{2 k_B T}{\pi^{3/2} f}\frac{1}{Y w (1-\sigma^2)} \phi_{sub},
\end{equation}
where  $\sigma$ is the Poisson ratio, and $\phi_{\rm{sub}}$ is the substrate loss angle. For the parameters considered, substrate thermal noise is expected to make a sub-dominant contribution to the total noise.

Current state-of-the-art optical cavities show many more mechanical resonances in the frequency band of interest ~\cite{chalermsongsak2014broadband} than the simplified model used in Fig.~\ref{Fig:Noise}.
These resonances could masquerade as a VULF DM signal; additionally they will reduce the sensitivity to the VULF DM signal in narrow bands distributed throughout the frequency range of interest. Fortunately, it is straightforward to solve both problems. The expected VULF DM signal is narrow-band, with an effective $Q$ factor corresponding to approximately $10^{6}$ and of strongly asymmetric shape~\cite{Derevianko2016a}; this is much narrower than any expected mechanical resonance. Moreover, by changing the temperature of the Fabry-Perot, the frequency of mechanical resonances will shift, while any VULF signal will not. Not only will this allow a VULF signal to be distinguished from a mechanical resonance, but it will allow the ``baseline'' sensitivity limits of Fig. 2 to be achieved across the entire bandwidth.

\begin{figure}[!t]
\begin{center}
\includegraphics[width=1.0\columnwidth]{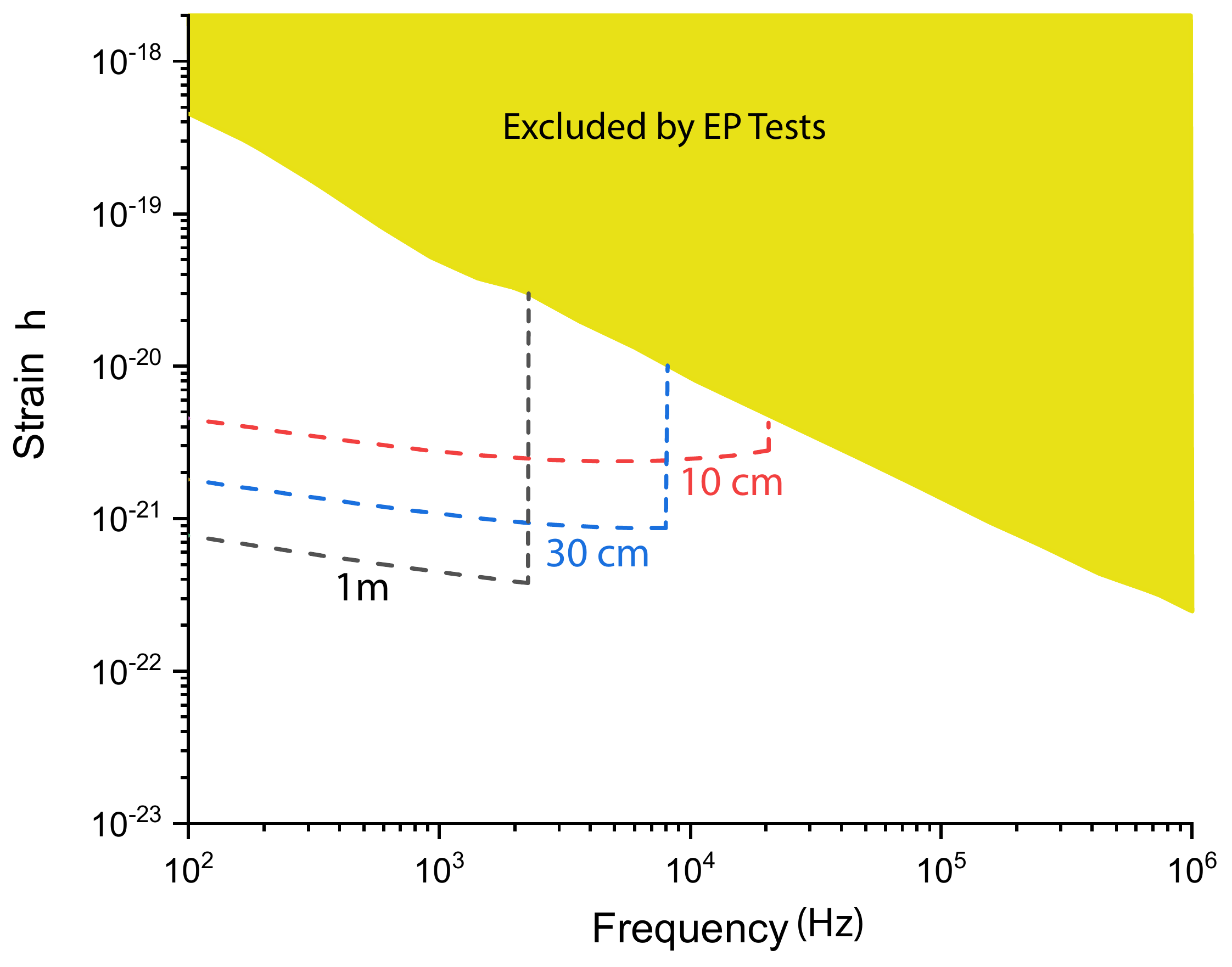}
\caption{Strain sensitivity of optical cavity limited by thermal noise for three cavity lengths as a function of VULF Compton frequency. Here we assume a total integration time of $10^7$ seconds, with an improvement scaling with the averaging time $\tau$ as $\tau^{-1/2}$ up to the coherence time of the DM field ($\sim 10^6$ oscillations), and improving as $\tau^{-1/4}$ thereafter.  Bounds from equivalence principle (EP) tests are shown as shaded yellow region \cite{EPtests1,EPtests2}.
\label{strain}}
\end{center}
\end{figure}

\begin{figure}[!t]
\begin{center}
\includegraphics[width=1.0\columnwidth]{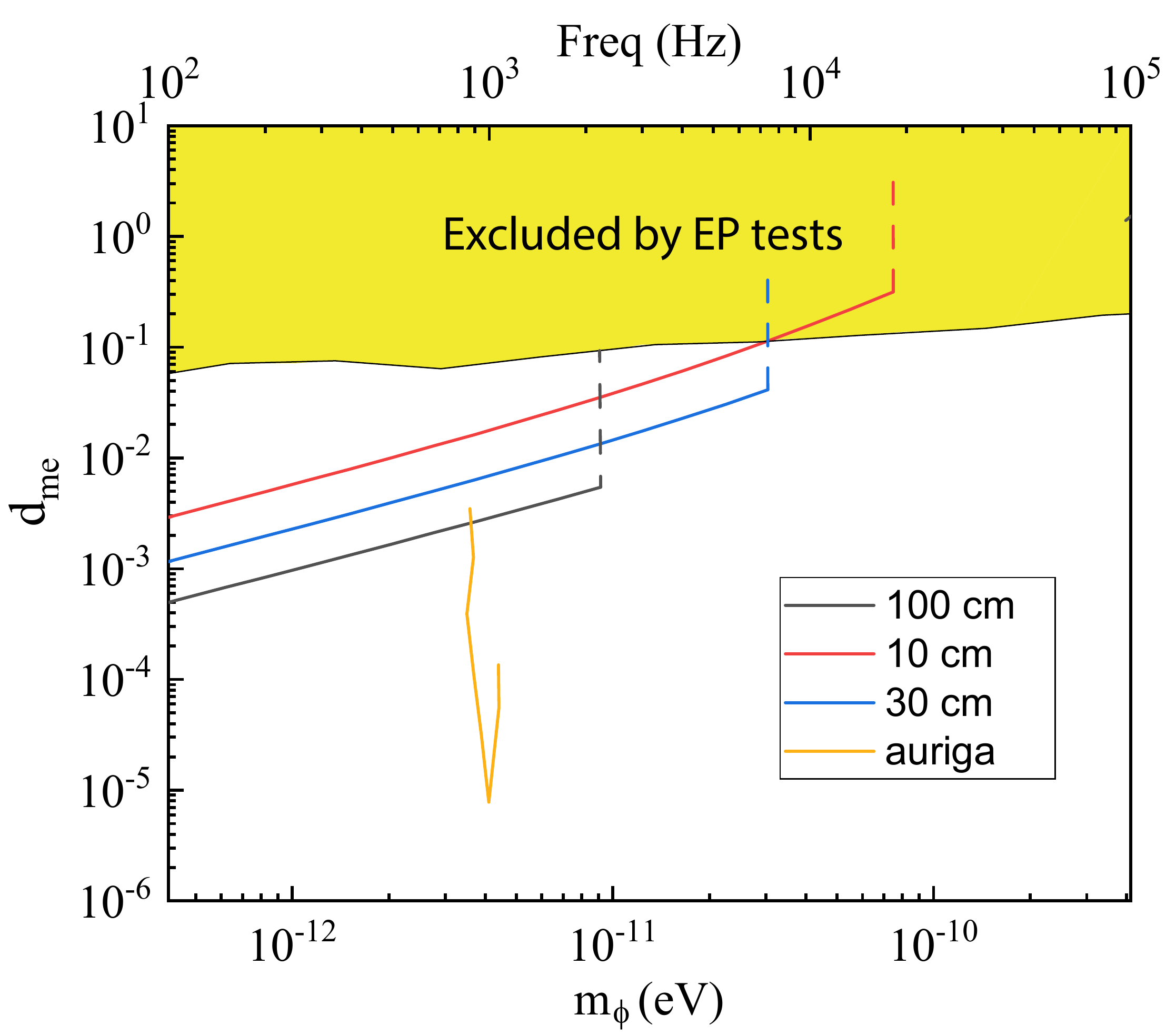}
\caption{Search reach for an optical cavity of length 10, 30, and 100 cm assuming thermal-noise limited sensitivity along with current experimental bounds from equivalence principle (EP) and fifth force tests \cite{EPtests1,EPtests2} as well as the limits derived from the narrow-band AURIGA gravitational wave detector \cite{AURIGA,Arvanitaki2016_SoundDM}.
\label{fig:dmeplot}}
\end{center}
\end{figure}

\emph{Results---}
Assuming we are limited by thermal noise as indicated in Fig.~\ref{Fig:Noise}, we plot the search reach for ultra-light scalar DM in terms of the strain $h$ and the constraints on the electron coupling $d_{m_e}$, along with bounds from equivalence principle tests and other experimental data in Figs.~\ref{strain} and \ref{fig:dmeplot}, respectively.
In general, the proposed techniques is sensitive to the combination $|d_e +d_{m_e}|$; in Fig.~\ref{fig:dmeplot} we assumed that $d_e$ coupling is negligible.  
For simplicity, we terminated the upper-frequency limit of the search range at the mechanical resonance frequency of the spacer.
Several orders of magnitude of improvement beyond the limits imposed by precision equivalence principle and fifth-force tests \cite{EPtests1,EPtests2} are possible at frequencies between 100 Hz and 10 kHz, depending on the length of the resonator. We also indicate the limits imposed by an analysis of the narrow-band AURIGA gravitational wave detector \cite{Arvanitaki2016_SoundDM,AURIGA}. 

\emph{Discussion ---}
A possible extension of the proposed technique could involve operating a network of spatially separated pairs of such cavities. Here we adopt general discussion~\cite{Derevianko2016a} to our specific case.  For the VULF Compton frequency range considered here, even an intercontinental  network is within the VULF coherence length. Then the sensitivity of the network improves as $\sqrt{N}$ with the number of nodes $N$. In the event of  positive VULF signal discovery, a network would provide a crucial DM signature as it would allow to measure the average direction for the incident DM waves. This direction, according to the standard halo model, should coincide  with the direction from the Cygnus constellation. 
As high-finesse optical  cavities are commonly in use in optical standards laboratories worldwide, the implementation of such a network may be relatively low-cost when compared with other proposed cryogenic strain-based sensing approaches \cite{Arvanitaki2016_SoundDM}.

While optical cavities with suspended mirrors are currently not commonly found in optical standards laboratories, it is common to find multiple rigid cavities of differing length, which would be sufficient to implement the proposed experiment. The DM-induced length fluctuations are suppressed above the resonant frequency of the cavity's mechanical spacer. Comparing two rigid cavities of different lengths allows the detection of a differential signal due to DM induced strain over the frequency band between the resonant frequencies of the two cavities, with the sensitivity of the shorter cavity.


Another extension of the current proposal is a search for ``clumpy'' DM composed of macroscopic  objects, such as Q-balls~\cite{KusSte01}, that lead to transient variations of fundamental constants~\cite{Roberts2018}. For these models, one may either rely on the annual variation in the measured noise non-Gaussianity for a single setup~\cite{Roberts2018} or on measuring correlated propagation of  variation in fundamental constants at $\sim 300 \, \mathrm{km/s}$ galactic velocities through the network~\cite{DerPos14}.    

Finally, the speed of sound in fused silica is $3750$ m/s, limiting the bandwidth of the detector to $3.7$ kHz for a $1$ meter cavity or 12.5 kHz for a 0.3 m cavity. For extending the search range to higher Compton frequency, silicon has a higher sound speed of $5800$ m/s. Cryogenic silicon cavities are also a promising route to improved thermal noise performance, as recently demonstrated by the atomic clock community \cite{juncryoclock}, and could extend the sensitivity to VULF DM by an order of magnitude.

\emph{Acknowledgements}
We would like to thank P. Hamilton, H. Muller, D. Schlippert, M. Arvanitaki, S. Dimopoulos, G. Ranjit, M. Baryakhtar, J. Huang, E. Rasel, B. Roberts, and R. Walsworth for discussions.
This work was supported in part by the US National Science Foundation and by the National Key Research Program of China under Grant 2016YFA0302002, and the Strategic Priority Research Program of the Chinese Academy of Sciences under Grant XDB21010100.

\bibliography{library-apd,library_jw,library_geraci} 
\end{document}